\documentclass[a4paper,11pt]{article}
\usepackage{pos}

\usepackage[utf8]{inputenc}
\usepackage{graphicx}
\usepackage{xspace}
\usepackage{subcaption}

\newcommand{\nnlojet}{NNLO\scalebox{0.8}{JET}\xspace}

\title{NNLO Photon Production with Realistic Photon Isolation}
\author[a,b]{X. Chen}
\author[c]{T. Gehrmann}
\author[d]{E.W.N. Glover}
\author[e]{M. H\"ofer}
\author[f]{A. Huss}
\author*[c]{R. Sch\"urmann}

\affiliation[a]{Institute for Theoretical Physics, Karlsruhe Institute of Technology,\\ 76131 Karlsruhe, Germany}
\affiliation[b]{Institute for Astroparticle Physics, Karlsruhe Institute of Technology,\\ 76344 Eggenstein-Leopoldshafen, Germany}
\affiliation[c]{Physik-Institut, Universit\"at Z\"urich,\\ Winterthurerstrasse 190, 8057 Z\"urich, Switzerland}
\affiliation[d]{Institute for Particle Physics Phenomenology, Department of Physics,\\ University of Durham, Durham, DH1 3LE, UK}
\affiliation[e]{Ludwig-Maximilians-Universit\"at M\"unchen, Fakult\"at f\"ur Physik, \\  Arnold Sommerfeld Center for Theoretical Physics, D-80333 M\"unchen, Germany}
\affiliation[f]{Theoretical Physics Department, CERN,\\ 1211 Geneva 23, Switzerland}

\emailAdd{xuan.chen@kit.edu}
\emailAdd{thomas.gehrmann@uzh.ch}
\emailAdd{e.w.n.glover@durham.ac.uk}
\emailAdd{m.hoefer@physik.uni-muenchen.de}
\emailAdd{alexander.huss@cern.ch}
\emailAdd{robins@physik.uzh.ch}

\abstract{ Isolated photon production at hadron colliders proceeds via direct production and fragmentation processes. Theory predictions for the isolated photon and photon-plus-jet cross section often impose idealised photon isolation criteria, eliminating the fragmentation contribution and introducing a systematic uncertainty in the comparison to data. We present NNLO predictions for the photon-plus-jet cross section with the experimental isolation including both, direct and fragmentation contributions. Predictions with two different parton-to-photon fragmentation functions are compared, allowing for an estimation of the uncertainty stemming from the only loosely constrained photon fragmentation functions.}

\FullConference{%
  Loops and Legs in Quantum Field Theory - LL2022,\\
  25-30 April, 2022\\
  Ettal, Germany
}

\begin{document}
\maketitle

\section{Introduction}

Photons are very frequently produced in hadronic collisions. Compared to the production of final-state partons, which undergo a hadronisation processes before they are detected, final-state photons can directly be measured by means of their distinct signature in the electromagnetic calorimeter. Therefore, measurements of photons in hadronic collisions give clean probes of the underlying hard scattering event and are of particular importance for testing our understanding of perturbative QCD. A complication in the analysis of photon cross sections arises from photons not produced in the hard scattering event. These non-direct photons can either stem from fragmentation processes, i.e.\ the fragmentation of a jet into a photon during hadronisation described by non-perturbative parton-to-photon fragmentation functions~\cite{Koller:1978kq,Laermann:1982jr}, or hadron decays, e.g.\ $\pi^0 \to \gamma \gamma$.  Although these non-direct photons typically have smaller transverse momenta than direct photons, they also yield a large background in the fiducial phase space volume. This complication is overcome in experimental analyses by imposing a photon isolation which aims to disentangle direct from non-direct photons. 

All analyses of the inclusive isolated photon~\cite{D0:2005ofv,CDF:2017cuc,ATLAS:2016fta,ATLAS:2017nah,ATLAS:2019buk,CMS:2018qao} and photon-plus-jet cross section~\cite{D0:2008chx,D0:2013lra,ATLAS:2017xqp,CMS:2018qao,CMS:2019jlq} in hadronic collisions have applied a cone based photon isolation in which the hadronic energy in a fixed-size cone around the photon must not exceed a certain threshold value.  The latest measurements of non-isolated photon production have been performed at LEP~\cite{Buskulic:1995au,Ackerstaff:1997nha}. 

Until recently, all next-to-next-to leading order (NNLO) QCD calculations for isolated photon production~\cite{Campbell_2017,Campbell:2017dqk,Chen_2020} employed idealised isolation procedures~\cite{Frixione:1998jh,Siegert:2016bre}, eliminating all hadronic radiation in the exact photon direction. Using these types of idealised isolations, the theory predictions neither have to handle singular collinear parton-photon configurations, nor do they have to include photon fragmentation processes. 
While simplifying the theoretical calculation, the use of these isolations limits the comparison between theory predictions and experimental data: idealised isolations rely on an empirical tuning of isolation parameters to approximate the experimental isolation and therefore introduce an additional source of uncertainty in the theoretical calculation. 
This drawback of previous state-of-the-art predictions has recently been overcome by a new NNLO calculation for the photon production cross section using the experimental isolation criterion~\cite{Chen:2022gpk}, which builds on the implementation of photon fragmentation processes in the antenna subtraction formalism~\cite{Gehrmann:2022cih}. 

Calculations using experimental photon isolations depend on the photon fragmentation functions. These functions obey inhomogeneous DGLAP evolution equations and their unknown boundary distributions must be modelled or extracted from experimental data. Since only sparse data sets suitable for the extraction of the fragmentation functions are available, most fragmentation function sets model this boundary condition~\cite{Owens:1986mp,Gluck:1992zx,Bourhis:1997yu}. 
So far, only non-isolated photon data~\cite{Buskulic:1995au,Ackerstaff:1997nha} were used for an extraction of the fragmentation functions from measurements~\cite{Gehrmann-DeRidder:1997fom}.

In this study, we compare our NNLO QCD predictions for the photon-plus-jet cross section to ATLAS data~\cite{ATLAS:2017xqp}. In the theory predictions the experimental isolation is used, avoiding any mismatch in the theory-data comparison. By computing the theory predictions with two different parton-to-photon fragmentation function sets, we estimate the uncertainty stemming from the only loosely constrained photon fragmentation functions.

\section{Photon Isolation and Fragmentation}

Direct photons are typically well separated from hadronic debris while most of the fragmentation photons and photons from hadron decays are contaminated by hadronic radiation. Therefore, a way to reduce the contribution from these non-direct photons is to impose a photon isolation. Isolation criteria allow only limited amount of hadronic energy in the vicinity of the photon. If the hadronic energy accompanying the photon exceeds a certain threshold value $E_T^{\mathrm{max}}$, the event is discarded.  

All experimental analyses to date use a fixed cone isolation in which the integrated hadronic energy inside a cone of radius $R=\sqrt{(\Delta \eta)^2 + (\Delta \phi)^2}$ in pseudorapidity $\eta$ and azimuthal angle $\phi$ around the photon candidate direction must remain below the maximal value $E_T^{\mathrm{max}}$. This threshold is commonly parametrised as
\begin{equation}
E_T^{\mathrm{max}} = \varepsilon \, p_T^{\gamma} + E_T^{\mathrm{thres}} \, ,
\label{eq:Etmaxfixed}
\end{equation}
where $p_T^{\gamma}$ denotes the transverse momentum of the photon candidate. The fixed cone isolation is characterised by the three parameters $(R, \varepsilon, \, E_T^{\mathrm{thres}} )$. It effectively eliminates photons from hadron decays but cannot fully suppress photons from fragmentation processes.

Theory predictions using the fixed cone isolation consequently depend on the parton-to-photon fragmentation functions $D_{p \to \gamma}(z,\mu_A^2)$, where $z$ is the longitudinal momentum fraction passed from parton $p$ to the photon and $\mu_A$ denotes the fragmentation scale at which the fragmentation functions are mass-factorised.
The photon fragmentation functions fulfil inhomogeneous DGLAP evolution equations. The boundary condition of the evolution cannot be calculated in a perturbative manner but must be modelled or extracted from data. In this work we focus on two determinations of the fragmentation functions:  the BFG set and the ALEPH parametrisation.

The most recent fragmentation function set is the BFG set~\cite{Bourhis:1997yu}, which incorporates the solution of the evolution equation at next-to-leading-logarithmic accuracy. The non-perturbative boundary condition is modelled by imposing an ansatz in which the parton-to-photon fragmentation functions are expressed as a superposition of parton-to-meson fragmentation functions. The BFG parametrisation consists of two sets of fragmentation functions, denoted by BFGI and BFGII. In our numerical study we use the BFGII set that describes flavour sensitive quark-to-photon fragmentation functions and also includes a gluon-to-photon fragmentation function. 

The ALEPH collaboration measured the quark-to-photon fragmentation function in events with two jets of which one carries an electromagnetic energy fraction $z>0.7$~\cite{Buskulic:1995au}. By comparing the measured distribution to the theoretical calculation of the photon-plus-jet cross section at LO~\cite{Glover:1993xc}, the LO quark-to-photon fragmentation function 
\begin{equation}
D_{q \to \gamma}(z, \mu_A^2) =  \frac{\alpha Q_q^2}{2\pi} p^{(0)}_{\gamma q}(z) \log\left(\mu_A^2/\mu_0^2 \right)+ \frac{\alpha Q_q^2}{2\pi} \left( - p^{(0)}_{\gamma q}(z) \log (1-z)^2 - 13.26 \right) 
\label{eq:ALEPH_FF}
\end{equation}
has been extracted, where the starting scale is $\mu_0=0.14 \, \mathrm{GeV}$, $Q_q$ is the charge of quark $q$ and $p^{(0)}_{\gamma q}$ denotes the leading order quark-to-photon spitting function. Equation~\eqref{eq:ALEPH_FF} is the fixed order solution of the DGLAP equation at $\mathcal{O}(\alpha)$. The quark-to-photon fragmentation function at $\mathcal{O}(\alpha \alpha_s)$ has been determined in~\cite{Gehrmann-DeRidder:1997fom}. The ALEPH parametrisation does not describe gluon-to-photon fragmentation processes.

\section{Implementation}

Our numerical results for the photon production cross section are obtained using the parton-level Monte Carlo event generator \nnlojet, which implements the antenna subtraction formalism~\cite{GehrmannDeRidder:2005cm,Daleo:2006xa,Currie:2013vh} at NNLO. 
All matrix elements required for the computation of NNLO QCD corrections to the photon production cross section are known in analytic form~\cite{Anastasiou:2001sv,Bern:2003ck,Bern:1994fz,Signer:1995np,Signer:1995a,DelDuca:1999pa} and have been implemented in \nnlojet. 

Using the experimental fixed cone isolation, our predictions have to account for both: photon fragmentation processes and singular parton-photon collinear configurations. 
These configurations have to be subtracted while retaining the information on the photon in the collinear cluster and upon integration are absorbed into the fragmentation functions by means of collinear factorisation~\cite{Koller:1978kq,Laermann:1982jr}. Within antenna subtraction, new phase space factorisations and new fragmentation antenna functions were introduced to handle photon fragmentation processes at NNLO~\cite{Gehrmann:2022cih}. As far as the subtraction of genuine QCD infrared singularities is concerned, our implementation builds on the subtraction terms used in a previous calculation~\cite{Chen_2020} with \nnlojet. Originally constructed for an idealised isolation, these subtraction terms can be combined with additional fragmentation subtraction terms for predictions employing a fixed cone isolation.

In our calculation the central values for the renormalisation scale $\mu_R$ and the initial-state factorisation scale $\mu_F$ are set to the transverse momentum of the photon, i.e.\ $\mu_R = \mu_F = p_T^{\gamma}$, and the fragmentation scale $\mu_A$ to the maximal invariant mass inside the isolation cone, i.e.\
\begin{equation}
\mu_A =m_{\mathrm{cone}} = \sqrt{E_T^{\mathrm{max}} \, p_T^{\gamma} \, R^2} + \mathcal{O}(R^3) \, .
\label{eq:muA_choice}
\end{equation}
Here, $R$ denotes the radius of the isolation cone and $E_T^{\mathrm{max}}$ is the maximal hadronic energy inside the cone~\eqref{eq:Etmaxfixed}.
This particular choice of the fragmentation scale reflects that additional parton radiation accompanying the fragmentation process is restricted by the photon isolation criterion. Since experimental analyses commonly impose tight isolation criteria, the numerical values for the fragmentation scale~\eqref{eq:muA_choice} are small compared to the values of the renormalisation and factorisation scale. We determine the scale uncertainty by varying the three scales by a factor of two around their central value, i.e.\
\begin{equation}
\mu_R = a \, p_T^{\gamma} \, , \quad \mu_F = b \, p_T^{\gamma} \,  , \quad \mu_A = c \, m_{\mathrm{cone}} \, , \quad a,b,c \in \big\{ \tfrac{1}{2}, 2 \big\} \, ,
\label{eq:default_scales}
\end{equation}
and we only keep combinations with $ 1/2 < a/b , b/c , a/c < 2$. The scale uncertainty band is given by the envelope of these 15 combinations. In our comparison to data from the ATLAS 13\,TeV photon-plus-jet study~\cite{ATLAS:2017xqp} we use the BFGII fragmentation function set~\cite{Bourhis:1997yu} and choose the NNPDF4.0 PDF set~\cite{NNPDF:2021njg}. 

The photon production cross section is composed of a direct and a fragmentation contribution. The different orders in the perturbative expansion take the form
\begin{eqnarray}
\mathrm{d} \hat{\sigma}^{\gamma+ X,{\rm LO}} &= &\mathrm{d} \hat{\sigma}_{\rm dir}^{\rm LO} \, ,\nonumber \\
\mathrm{d} \hat{\sigma}^{\gamma+ X,{\rm NLO}} &= & \mathrm{d} \hat{\sigma}_{\rm dir}^{\rm NLO} +  \mathrm{d} \hat{\sigma}^{\rm NLO}_{\rm frag}  \, , \nonumber \\
\mathrm{d} \hat{\sigma}^{\gamma+ X,{\rm NNLO}} &= & \mathrm{d} \hat{\sigma}_{\rm dir}^{\rm NNLO} + \mathrm{d} \hat{\sigma}_{\rm frag}^{\rm NNLO} \, ,
\label{eq:sigma}
\end{eqnarray}
where we assigned an $\mathcal{O}(\alpha)$ power counting to the fragmentation functions. This counting of powers is warranted by our choice of the fragmentation scale and the overall numerically small contribution from fragmentation processes.

\section{Numerical Results}

We compare our NNLO cross section to data of the ATLAS 13\,TeV photon-plus-jet measurement~\cite{ATLAS:2017xqp}. In the experimental analysis only photon candidates with $p_T^{\gamma} > 125\, \mathrm{GeV}$ and  $|\eta_{\gamma}|<1.37$ or $1.56 < |\eta_{\gamma}|<2.37$ are considered. Moreover, a fixed cone isolation with parameters
\begin{equation}
R=0.4 \, , \quad \varepsilon = 0.0042 \, ,\quad E_T^{\mathrm{thres}} = 10 \, \mathrm{GeV} \, ,
\end{equation} 
is imposed. Jets are reconstructed using an anti-$k_T$ algorithm with a cone radius $R_{\mathrm{jet}} = 0.4$. The reconstructed jet must be separated from the photon candidate by $\Delta R_{\gamma \mathrm{jet}} > 0.8$ and it has to fulfil $|y_{\mathrm{jet}}|<2.37$ and $p_T^{\mathrm{jet}} > 100 \, \mathrm{GeV}$. 

\begin{figure}[!t]
\centering
\begin{subfigure}[b]{0.496\textwidth}
\centering
\includegraphics[width=\textwidth]{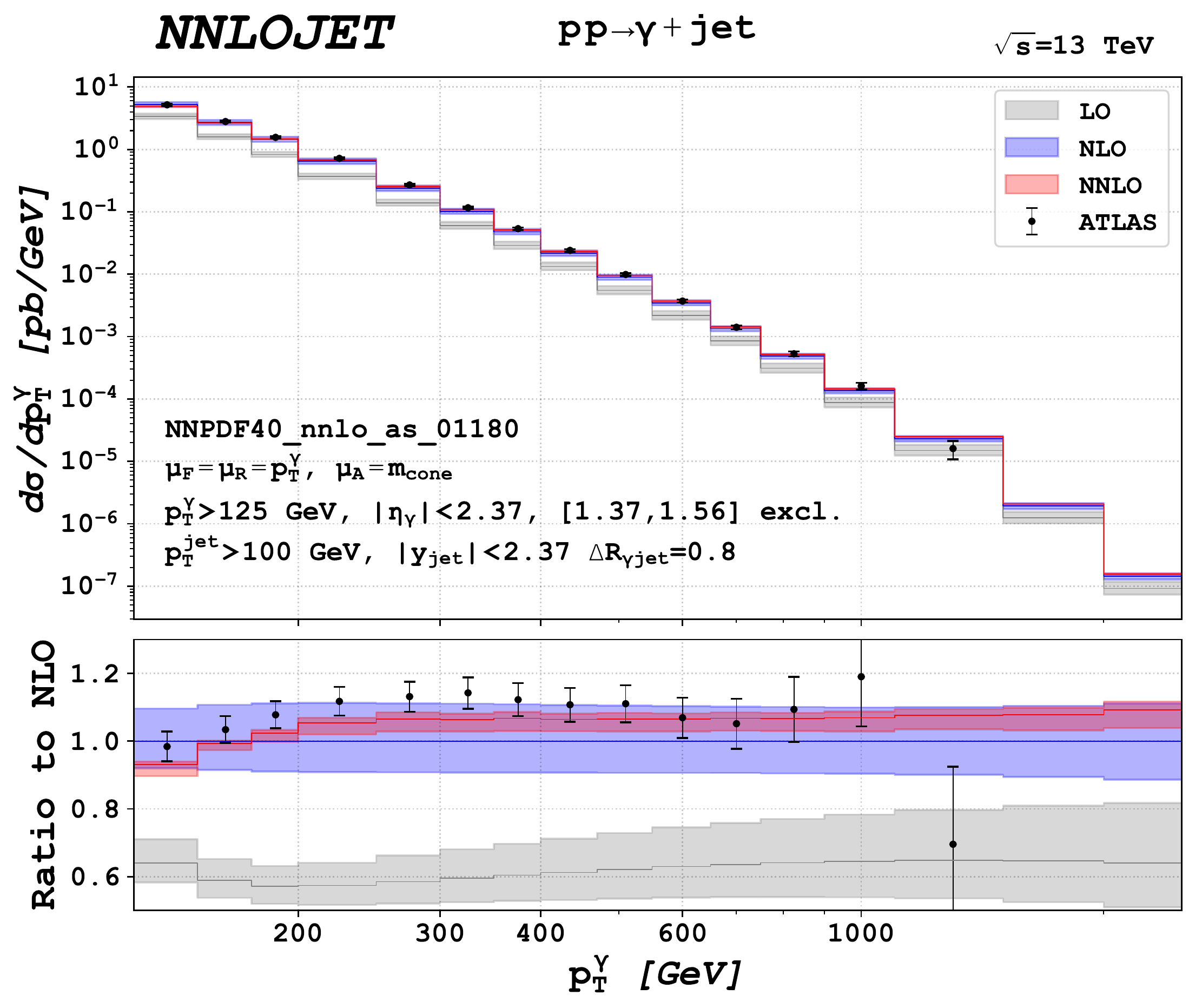}
\end{subfigure}
\hfill
\begin{subfigure}[b]{0.496\textwidth}
\centering
\includegraphics[width=\textwidth]{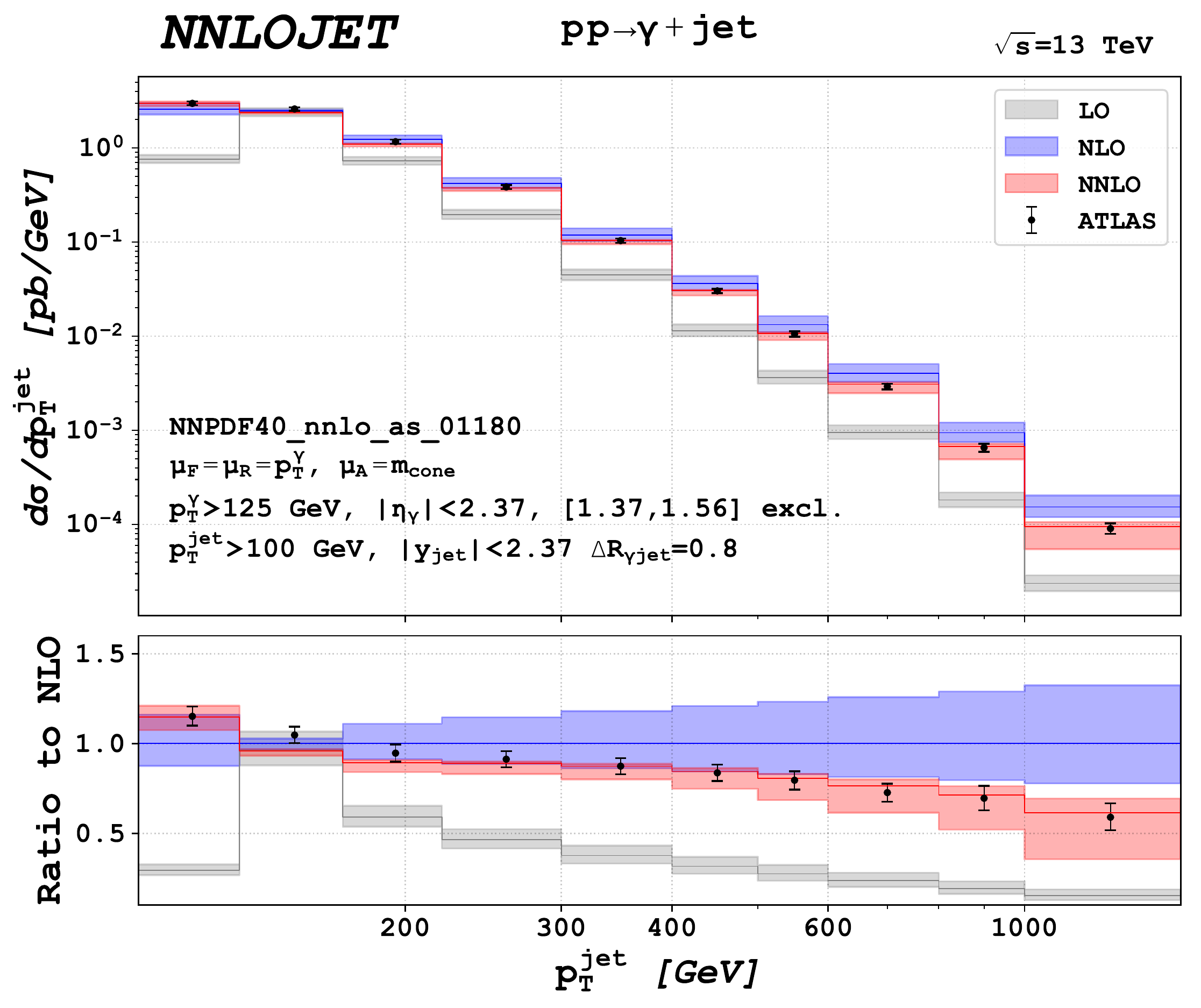}
\end{subfigure}
\caption{Predictions for the photon transverse momentum distribution (left) and jet transverse momentum distribution (right) at LO, NLO, and NNLO. The predictions are compared to data from the ATLAS 13\,TeV photon-plus-jet measurement~\protect\cite{ATLAS:2017xqp}.}
\label{fig:ATLAS_ptgam_ptjet}
\end{figure}

The comparison of our predictions to the measured photon and jet transverse momentum distribution is shown in Figure~\ref{fig:ATLAS_ptgam_ptjet}.
The inclusion of NNLO corrections strongly impacts the shape of the two distributions compared to the NLO approximation of the cross section. At small photon transverse momenta, the NNLO corrections are negative and reduce the cross section by -7\,\%, while for $p_T^{\gamma} \gtrsim 300 \, \mathrm{GeV}$ the NNLO corrections uniformly increase the cross section by 7\,\%.  
In the first bin of the jet transverse momentum distribution the NNLO-to-NLO K-factor is 1.15. Towards larger transverse momenta the K-factor decreases and at $p_T^{\mathrm{jet}} \sim 1 \, \mathrm{TeV}$ it amounts to only 0.60. This kinematic regime is dominated by events with two highly energetic jets accompanied by a relatively soft photon. For these events our NNLO predictions are effectively of NLO accuracy, causing this particularly small NNLO-to-NLO K-factor in the tail of the $p_T^{\mathrm{jet}}$ distribution.

Over a wide kinematic range the scale uncertainty at NNLO is small and does not exceed 5-7\,\%. Only in the low and high $p_T^{\mathrm{jet}}$ regime the uncertainty remains larger. Due to the asymmetric $p_T$ cuts on the photon and the jet, the first bin of the jet transverse momentum distribution is effectively populated only from NLO onwards,
resulting in a 12\,\% NNLO scale uncertainty in this bin. In the tail of the $p_T^{\mathrm{jet}}$ distribution our particular scale choice of $\mu_R = \mu_F = p_T^{\gamma}$ yields large scale ratios of $p_T^{\gamma} / p_T^{\mathrm{jet}}$ for a majority of the events. In this regime the scale uncertainty is $\sim$ 55\,\% .

Comparing our predictions to the measured distributions, we observe perfect agreement in the shapes of the distributions once NNLO corrections are taken into account. For the jet transverse momentum distribution data and theory predictions coincide within their respective uncertainty over almost the entire kinematic range. It is only in the second bin of this distribution that the theory predictions slightly undershoot the data. For $p_T^{\gamma} \lesssim 350 \, \mathrm{GeV}$ the theory predictions systematically fall below the measured distribution and theory predictions and data are not fully compatible within their uncertainties in the regime. At larger photon transverse momenta, theory predictions and data are compatible. 

\section{Uncertainty from the Choice of Fragmentation Functions}

The theoretical uncertainties obtained from the 15-point scale variation take into account missing higher order effects but do not include the uncertainty originating from the fragmentation functions. 
To get an estimate of the uncertainty stemming from these poorly constrained functions, we recomputed the NNLO theory predictions for the photon-plus-jet cross section with the LO ALEPH fragmentation functions~\eqref{eq:ALEPH_FF} and compared these predictions with our benchmark predictions using the BFGII set. The kinematic set-up of the ATLAS 13\,TeV photon-plus-jet study~\cite{ATLAS:2017xqp} is used and the scales are set according to~\eqref{eq:default_scales}. 
We note that the use of the LO ALEPH parametrisation instead of the NLO parametrisation is not fully consistent in NNLO predictions. Only the NLO fragmentation functions incorporate the exact solution of the evolution equation at $\mathcal{O}(\alpha \alpha_s)$. Notwithstanding, our comparison employing the LO ALEPH parametrisation enables us to give an estimate of the overall uncertainty from the choice of fragmentation functions.

\begin{table}[!t]
\centering
\begin{tabular}{|@{\hspace{0.8em}}l@{\hspace{0.8em}}|r|r|@{\hspace{0.0em}}c@{\hspace{0.8em}}|}
\hline
\multicolumn{1}{|c|}{}                   & \multicolumn{1}{c|}{BFGII}             & \multicolumn{1}{c|}{ALEPH}            & $\, \, \, \sigma^{\mathrm{ALEPH}}/\sigma^{\mathrm{BFGII}}$   \\ \hline
$\sigma^{\mathrm{NLO}}_{\mathrm{frag}}$  & $(8.525 \pm 0.001) \, \mathrm{pb}$  & $(14.094 \pm 0.003) \, \mathrm{pb}$ & $1.653$ \\ \hline
$\sigma^{\mathrm{NNLO}}_{\mathrm{frag}}$ & $(11.61 \pm 0.01) \, \mathrm{pb}$ & $(18.02 \pm 0.02) \, \mathrm{pb}$   & $1.551$ \\ \hline
$\sigma^{\mathrm{NNLO}}$                 & $(284.3 \pm 0.2) \, \mathrm{pb}$    & $(290.7 \pm 0.2) \, \mathrm{pb}$    & $1.022$ \\ \hline
\end{tabular}
\caption{Comparison of predictions for the integrated photon-plus-jet cross section obtained with the BFGII set of fragmentation functions and the LO ALEPH fragmentation functions. $\sigma^{\mathrm{NNLO}}$ refers to the full NNLO cross section, taking into account the fragmentation and direct contribution.}
\label{tab:FF_compare_default}
\end{table}

Table~\ref{tab:FF_compare_default} compares the two predictions at the level of the integrated cross section. The ALEPH fragmentation functions yield a 65\,\% larger NLO fragmentation contribution compared to the BFGII set. Taking into account NNLO corrections to fragmentation processes, this discrepancy is reduced but it remains larger than~55\,\%. Fragmentation processes contribute only $\sim 5\,\%$ to the full NNLO cross section. Nevertheless, the very large uncertainty on this small contribution results in a $\sim 2 \,\%$ difference at the level of the full NNLO cross section. Given that the scale uncertainties of the NNLO cross section are only a few percent, this shows that the uncertainty from the choice of fragmentation functions is non-negligible also for tightly isolated cross sections.

\begin{figure}[!t]
\centering
\begin{subfigure}[b]{0.496\textwidth}
\centering
\includegraphics[width=\textwidth]{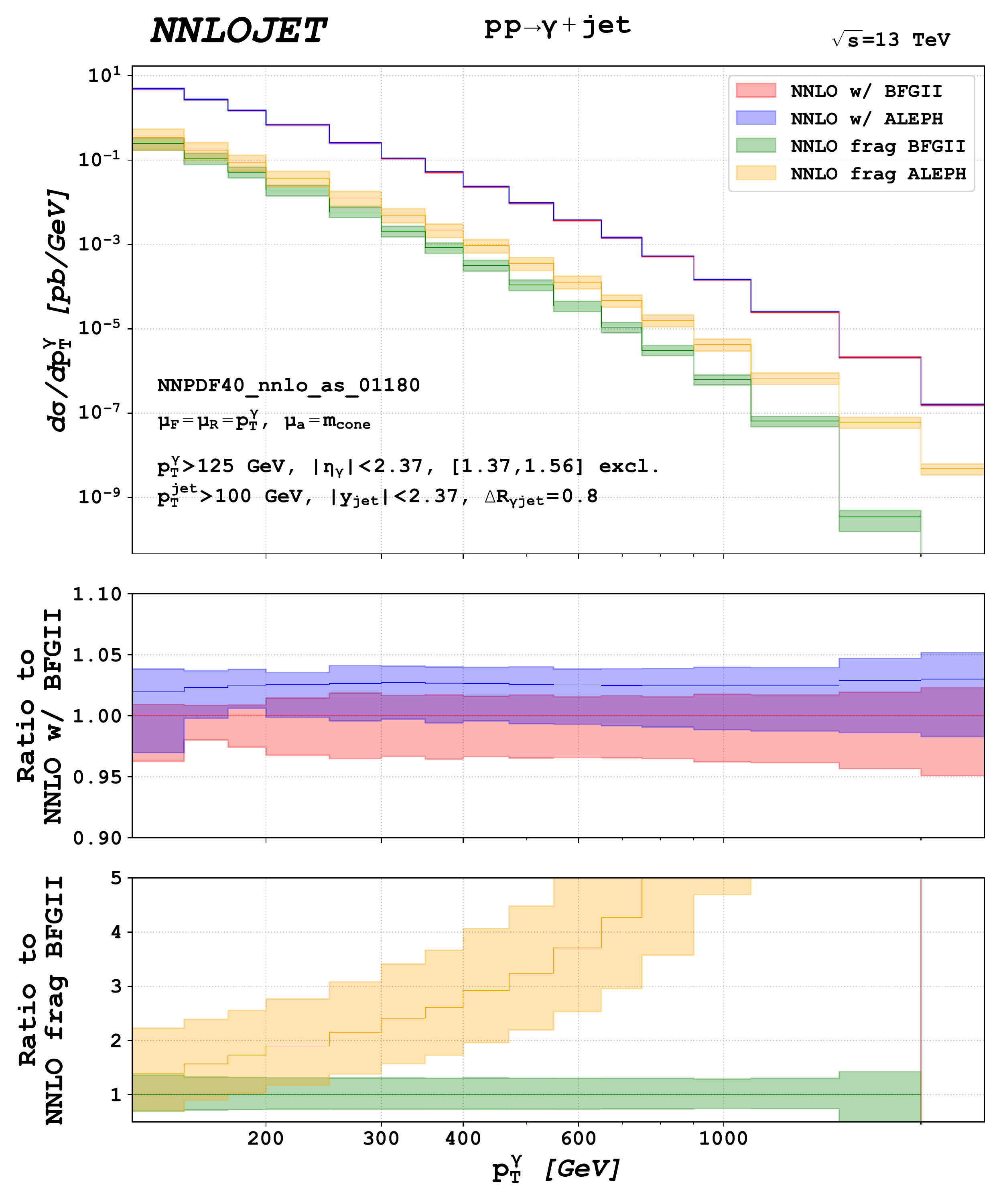}
\end{subfigure}
\hfill
\begin{subfigure}[b]{0.496\textwidth}
\centering
\includegraphics[width=\textwidth]{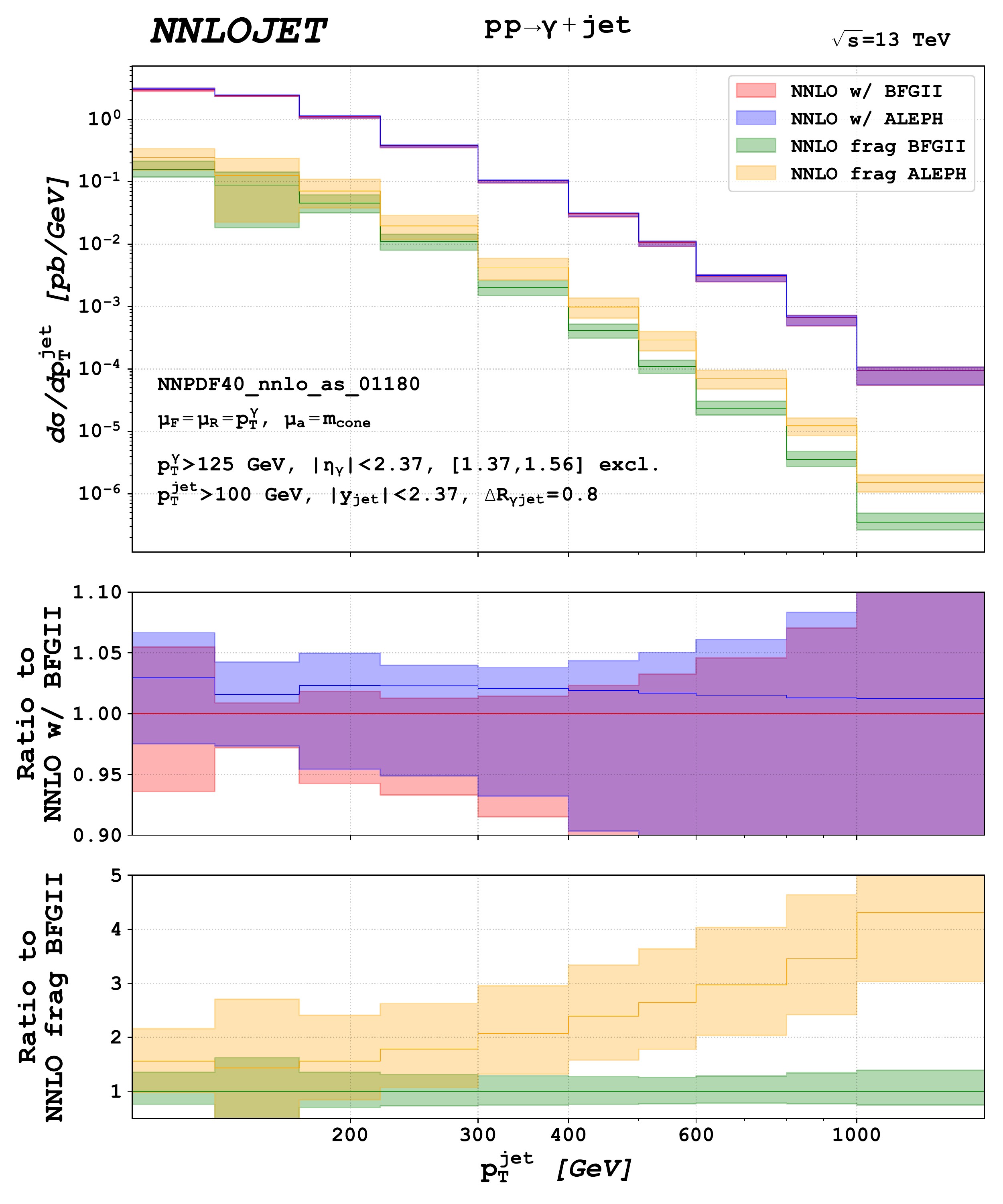}
\end{subfigure}
\caption{Predictions for the photon-plus-jet cross section differential in the photon transverse momentum (left) and the jet transverse momentum (right) obtained with the BFGII set and the LO ALEPH parametrisation of the fragmentation functions.}
\label{fig:comp_FF_tight}
\end{figure}

In Figure~\ref{fig:comp_FF_tight} the cross sections differential in the photon transverse momentum and jet transverse momentum are compared. 
For the photon transverse momentum distribution the ratio of the full NNLO cross sections obtained with the two fragmentation function sets is $\sim 1.02$ for small to mid $p_T^{\gamma}$ and slightly increases towards very large values of the photon transverse momentum, where it amounts to $\sim 1.03$. Comparing only the fragmentation contributions (lower panel in Figure~\ref{fig:comp_FF_tight}), a strong enhancement of the ALEPH cross section with respect to the BFGII contribution towards high $p_T^{\gamma}$ is observed. At very large photon transverse momenta, the fragmentation functions are probed in the very near vicinity of unity. In this regime the LO ALEPH fragmentation functions are particularly enhanced compared to the BFGII set due to the explicit logarithm $\log(1-z)^2$ in~\eqref{eq:ALEPH_FF}. The relative contribution from fragmentation processes in this kinematic regime is small compared to the direct contribution so that this enhancement does not strongly affect the full NNLO cross section.

For the jet transverse momentum distribution the ratio of the full NNLO cross section obtained with the two fragmentation function sets is decreasing with increasing $p_T^{\mathrm{jet}}$. In the first bin the ratio is $1.03$, while at $p_T^{\mathrm{jet}} \sim 1 \, \mathrm{TeV}$ it is only $1.01$. The ALEPH fragmentation contribution increases strongly compared to the BFGII contribution with increasing jet transverse momenta, reflecting that events with only one high-$p_T$ jet require a recoiling photon with very large transverse momentum. 

The comparison of the NNLO predictions for the photon-plus-jet cross section obtained with the two different fragmentation function sets reveals that the uncertainty stemming from the fragmentation functions is of the order of a few percent and therefore not negligible. 
For tight isolations, the fragmentation functions are probed in the close vicinity of unity. In this regime the BFGII fragmentation functions are almost unconstrained and yield very different results compared to the LO ALEPH parametrisation.

\section{Conclusion}

In this work we have presented NNLO QCD predictions for the photon-plus-jet cross section using the photon isolation condition applied in the experimental measurement. The predictions are compared to data from ATLAS~\cite{ATLAS:2017xqp} and we find very good agreement over a wide kinematic range. By employing the experimental isolation in our theory predictions, they receive contributions from photon fragmentation processes. We have quantified the uncertainty from the poorly constrained parton-to-photon fragmentation functions by comparing the BFGII set to the ALEPH parametrisation of the fragmentation functions. At the level of the NNLO cross section this uncertainty is a few percent. Since the scale uncertainty of the NNLO predictions is typically only $\sim 5\,\%$, the fragmentation functions contribute significantly to the overall theoretical uncertainty. This study indicates that for future precision phenomenology of processes with final-state photons a new assessment of the photon fragmentation functions is required.

\section*{Acknowledgement}
This work has received funding from the Swiss National Science Foundation (SNF) under contract 200020-204200, from the European Research Council (ERC) under the European Union's Horizon 2020 research and innovation programme grant agreement 101019620 (ERC Advanced Grant TOPUP), from the UK Science and Technology Facilities Council (STFC) through grant ST/T001011/1 and from the Deutsche Forschungsgemeinschaft (DFG, German Research Foundation) under grant 396021762-TRR 257. 

\bibliographystyle{JHEP}
\bibliography{LL2022_photons}

%\begin{thebibliography}{99}
%\bibitem{...}
%....
%
%\end{thebibliography}

\end{document}